\documentclass[preprint]{jpsj3}

\usepackage{graphicx}

\title{Analysis of Resonant Inelastic X-Ray Scattering in Stripe-Ordered Nickelate}

\author{Takuji \textsc{Nomura}\thanks{E-mail address: nomurat@spring8.or.jp} 
and Eiji Kaneshita$^1$}

\inst{
Quantum Beam Science Directorate, Japan Atomic Energy Agency, 
Sayo, Hyogo 679-5148, Japan \\
$^1$Sendai National College of Technology, 
4-16-1 Ayashichuo, Aoba-ku, Sendai 989-3128, Japan \\}

\recdate{\today}

\abst{
We analyze theoretically the resonant inelastic x-ray scattering (RIXS) 
at the Ni $K$ edge in the stripe-ordered state of La$_{2-x}$Sr$_x$NiO$_4$ at $x=1/3$. 
In the calculation of RIXS spectra, the stripe-ordered ground state is described 
within the Hartree-Fock approximation by using a realistic tight-binding model 
for Ni3$d\gamma$ and O2$p_{x, y}$ orbitals, 
and the electron correlations in the electronic excitation processes  
are taken into account within the random-phase approximation. 
The calculated RIXS spectrum shows a tail toward the low-energy region 
when the momentum transfer of photons equals the stripe vector ${\mib Q}$, 
being consistent with a recent experimental result. 
The origin of this anomalous momentum dependence of RIXS spectra 
is discussed microscopically.}

\kword
{
resonant inelastic x-ray scattering, stripe ordering, random-phase approximation, 
Hartree-Fock approximation, nickelate, La$_{5/3}$Sr$_{1/3}$NiO$_4$
}

\begin{document}
\maketitle

\section{Introduction}

Resonant inelastic x-ray scattering (RIXS) has been developed 
to be a powerful method for measuring elementary excitations 
in solids~\cite{Kotani2001, Kotani2005, Ament2011}. 
This is largely owing to both the recent advances in instrumentations 
and the achievement of highly brilliant lights obtained from advanced synchrotron facilities. 
In RIXS processes, the incident photon energy is tuned to match an absorption 
energy of one of the constituent elements. 
The electronic system of materials is resonantly excited by absorbing incident photons, 
and then after a short time (typically, of the order of femtoseconds), 
photons are emitted, whose momentum and energy generally differ from those 
of the incident photons. 
The electronic system remains an excited state still after the photon is emitted. 
The momentum change and energy loss of photons necessarily equal 
the momentum and energy spent for exciting the electronic system, 
due to the energy and momentum conservation laws. 
Therefore, one can get information on the electronic excitations of the system 
by measuring systematically the momentum change and energy loss of photons. 

Excitation processes involved in RIXS depend on the material 
and incident x-ray wavenumber utilized in the RIXS measurements. 
Among them, RIXS at the transition-metal $K$ edges has attracted much interest, 
partly because the RIXS spectra are naturally expected 
to reflect the strong electron correlations 
of $d$ electrons in transition-metal compounds, 
such as cuprates~\cite{Hill1998,Abbamonte1999,Hasan2000,Hasan2002,Kim2002,Kim2004,
Ishii2005a,Ishii2005b,Lu2005,Suga2005,Shukla2006,Collart2006,Ellis2008}, 
manganites~\cite{Inami2003,Grenier2005}, 
nickelates~\cite{Kao1996,Collart2006} etc. 
In the RIXS at the transition-metal $K$ edges, 
transition-metal $1s$ core electrons are resonantly excited 
to the transition-metal $4p$ unoccupied bands. 
In the intermediate states, the transition-metal $d$ electrons 
are excited to screen the created $1s$ core hole. 
Here we should note that the core hole potential, i.e., the Coulomb interaction 
between the transition-metal $1s$ and $d$ orbitals, plays an essential role. 
In the final state, the $4p$ electron is annihilated together with the $1s$ hole 
by emitting a photon, before the $d$ electrons decay from the excited state 
to the ground state. 
Overall, the momentum change and energy loss of photons 
are transferred indirectly to the excited $d$ electrons. 

The x-ray at the transition-metal $K$ edges is situated in the hard x-ray regime, 
and its wavenumber is appropriate for sweeping the whole Brillouin zone. 
Taking advantage of this point, momentum dependence of the RIXS spectra 
has been indeed observed in various transition-metal compounds, 
such as copper oxides~\cite{Hasan2000,Kim2002,Hasan2002,Kim2004,Ishii2005a,
Ishii2005b,Lu2005,Suga2005,Collart2006,Ellis2008}, 
NiO~\cite{Takahashi2007} etc. 
As a recent result from intensive experimental and theoretical researches, 
it has now become clear that the RIXS spectra at the transition-metal $K$ edges 
reflect the charge correlation function of the strongly correlated $d$ electrons~\cite{Nomura2005,Igarashi2006,Brink2006,Markiewicz2006,Ishii2005b,Kim2007,Markiewicz2008}. 
As well known, neutron scattering intensity is related to the spin correlation function. 
Thus, roughly speaking, RIXS in transition-metal compounds is to charge correlations 
what neutron scattering is to spin correlations. 
According to more recent researches, multi-magnon excitations have been observed 
at low-energy regions in insulating cuprates~\cite{Hill2008,Ellis2010}. 
Therefore RIXS may provide a promising way of studying not only charge excitations 
but also magnetic excitations of strongly correlated electrons 
in transition-metal compounds in the future. 

Stripe ordering has been one of the central issues 
in the physics of strongly correlated electron systems. 
Neutron scattering measurements have revealed that the holes and spins 
exhibit spatial disproportionation with a stripe form 
in several transition-metal oxides, e.g., 
La$_{2-x}$Sr$_x$NiO$_4$~\cite{Sachan1995,Tranquada1996,Lee1997,Yoshizawa2000}, 
La$_2$NiO$_{4+\delta}$~\cite{Tranquada1994, Tranquada1995a}, 
La$_{2-x}$(Ba, Sr)$_x$CuO$_4$ at $x=\frac{1}{8}$~\cite{Tranquada1995b}, etc. 
To try to elucidate relations to the high-$T_{\rm c}$ cuprate superconductivity, 
a lot of research works on stripe ordering have already been accumulated so far~\cite{Kivelson2003}. 
Recently, Wakimoto and collaborators reported RIXS in two stripe-ordered 214 compounds 
La$_{5/3}$Sr$_{1/3}$NiO$_4$ and $1/8$ doped La$_{2-x}$(Ba or Sr)$_x$CuO$_4$ 
at the Cu and Ni $K$ edges, respectively~\cite{Wakimoto2009}. 
They observed low-energy ($\simeq$ 1 eV) excitations with a momentum transfer 
corresponding to the charge stripe spatial period in the both compounds. 
The aim of our present work is to analyze this anomalous spectral feature, 
and discuss its microscopic origin for the case of La$_{5/3}$Sr$_{1/3}$NiO$_4$. 

The present article is constructed as follows. 
In \S~\ref{Sc:Formulation}, we present a model Hamiltonian, theoretical description 
of the stripe-ordered state, and the formula for RIXS intensity. 
We take the Hartree-Fock (HF) approximation for describing the stripe-ordered ground states, 
and the random-phase approximation (RPA) for taking account of the electron correlations
in the intermediate states of the excitation processes. 
In \S~\ref{Sc:Numerical}, numerical results of RIXS spectra are presented. 
There we present not only the results for stripe states (\S~\ref{Sc:Stripe})
but also for the undoped antiferromagnetic insulating state (\S~\ref{Sc:AF}). 
In \S~\ref{Sc:Conclusions}, the article is concluded with some discussions and remarks. 

\section{Formulation}
\label{Sc:Formulation}
\subsection{Model and Hartree-Fock approximation for stripe states}
\label{Sc:Model}

The electronic properties of La$_{2-x}$Sr$_x$NiO$_4$ are considered 
to be dominated by those of the NiO$_2$ layers. 
Therefore we can use two-dimensional tight-binding model 
for a single NiO layer to reproduce the in-plane electronic structure 
of La$_{2-x}$Sr$_x$NiO$_4$. 
Since only Ni$3d\gamma$ orbitals are the most essential 
among the five Ni$3d$ orbitals, we take the following four orbitals: 
Ni$3d_{x^2-y^2}$, Ni$3d_{3z^2-r^2}$, O2$p_x$ and O$2p_y$. 
Hereafter, we specify each of these four orbitals by index $\ell$: 
$\ell=1$ for Ni$3d_{x^2-y^2}$, $\ell=2$ for Ni$3d_{3z^2-r^2}$, 
$\ell=3$ for O$2p_x$, $\ell=4$ for O$2p_y$. 
The noninteracting part of the Hamiltonian is given by 
\begin{equation}
H_0= \sum_{i} \sum_{\ell\sigma} \varepsilon_{\ell} 
c^{\dag}_{i\ell\sigma} c_{i\ell\sigma} 
+ \sum_{i,j} \sum_{\ell\ell'\sigma} t_{\ell, \ell'}({\mib r}_{ij}) 
c^{\dag}_{i\ell\sigma} c_{j\ell'\sigma}, 
\label{eq:H_0}
\end{equation}
where ${\mib r}_{ij} = {\mib r}_i - {\mib r}_j$, $\ell$ and $\ell'$ are orbital indices, 
$c_{i\ell\sigma}$ and $c^{\dag}_{i\ell\sigma}$ are respectively the annihilation 
and creation operators for the electron with spin $\sigma$ on orbital $\ell$ at site $i$. 

For the one-particle energies, we take 
$\varepsilon_1=\varepsilon_2 \equiv \varepsilon_d=-9$ eV, 
$\varepsilon_3=\varepsilon_4 \equiv \varepsilon_p=0$ eV. 
For the hopping parameters we take 
$t_{1,3}(\frac{\hat{x}}{2})=-t_{1,4}(\frac{\hat{y}}{2}) \equiv t_{dp}=-1.2$ eV and 
$t_{2,3}(\frac{\hat{x}}{2})=t_{2,4}(\frac{\hat{y}}{2}) \equiv -t_{dp}/\sqrt{3}$ 
for nearest-neighbor Ni-O bonds, $t_{3,4}(\frac{\hat{x}+\hat{y}}{2})= 
-t_{3,4}(\frac{\hat{x}-\hat{y}}{2}) \equiv t_{pp}=0.6$ eV
for nearest-neighbor O-O bonds. ($\hat{x}$ and $\hat{y}$ are the unit lattice 
vectors connecting inplane nearest-neighbor Ni sites.) 
The hopping matrix $t_{\ell, \ell'}({\mib r})$ satisfies 
the relation $t_{\ell, \ell'}({\mib r}) = t_{\ell', \ell}(-{\mib r}) 
= s_{\ell} s_{\ell'} t_{\ell, \ell'}(-{\mib r}) $, 
where the factor $s_{\ell}$ equals $+1$ for $\ell = 1$ and $2$ (Ni$3d$ orbitals), 
and $-1$ for $\ell = 3$ and $4$ (O$2p$ orbitals). 
The Fourier transform of the above non-interacting Hamiltonian is 
given in the following form: 
\begin{equation}
H_0 = \sum_{{\mib k}} \sum_{\ell\ell'} \sum_{\sigma} \xi_{\ell\ell'}({\mib k}) 
c^{\dag}_{{\mib k}\ell\sigma} c_{{\mib k}\ell'\sigma}, 
\label{eq:H_0_2}
\end{equation}
with 
\begin{equation}
\xi_{\ell\ell'}({\mib k}) = \varepsilon_{\ell} \delta_{\ell\ell'} 
+ \sum_{\mib r} e^{- {\rm i} {\mib k} \cdot {\mib r}} t_{\ell, \ell'} ({\mib r}), 
\label{eq:xi}
\end{equation}

where ${\mib k}$ denotes momentum in the first Brillouin zone (BZ). 
For the convenience in the following, we introduce the annihilation operators 
$d$ and $p$ [creation operators $d^{\dag}$ and $p^{\dag}$] 
by $ (d_{i1\sigma}, d_{i2\sigma}, p_{i3\sigma}, p_{i4\sigma}) \equiv 
(c_{i1\sigma}, c_{i2\sigma}, c_{i3\sigma}, c_{i4\sigma})$ 
[ $ (d_{i1\sigma}^{\dag}, d_{i2\sigma}^{\dag}, p_{i3\sigma}^{\dag}, p_{i4\sigma}^{\dag}) \equiv 
(c_{i1\sigma}^{\dag}, c_{i2\sigma}^{\dag}, c_{i3\sigma}^{\dag}, c_{i4\sigma}^{\dag})$ ]. 
Namely, $d_{i\ell\sigma} \equiv c_{i\ell\sigma}$ when $i$ is on Ni site and $\ell=1$ or $2$, 
and $p_{i \ell \sigma} \equiv c_{i \ell \sigma}$ when $i$ is on O site and $\ell=3$ or $4$. 
For the momentum representation, we have 
$d_{{\mib k} \ell \sigma} \equiv c_{{\mib k} \ell \sigma}$ for $\ell=1$ and $2$, and 
$p_{{\mib k} \ell \sigma} \equiv c_{{\mib k} \ell \sigma}$ for $\ell=3$ and $4$. 

For the interacting part, we take the on-site Coulomb interaction at Ni sites: 
\begin{eqnarray}
H' &=& \frac{U}{2} \sum_{i} \sum_{\ell} \sum_{\sigma \neq \sigma'}
d_{i\ell\sigma}^{\dag} d_{i\ell\sigma'}^{\dag} d_{i\ell\sigma'} d_{i\ell\sigma} \nonumber\\
&& + \frac{U'}{2} \sum_{i} \sum_{\ell\neq\ell'} \sum_{\sigma,\sigma'}
d_{i\ell\sigma}^{\dag} d_{i\ell'\sigma'}^{\dag} d_{i\ell'\sigma'} d_{i\ell\sigma} \nonumber\\
&& + \frac{J}{2} \sum_{i} \sum_{\ell\neq\ell'} \sum_{\sigma,\sigma'}
d_{i\ell\sigma}^{\dag} d_{i\ell'\sigma'}^{\dag} d_{i\ell\sigma'} d_{i\ell'\sigma} \nonumber\\
&& + \frac{J'}{2} \sum_{i} \sum_{\ell\neq\ell'} \sum_{\sigma \neq \sigma'}
d_{i\ell\sigma}^{\dag} d_{i\ell\sigma'}^{\dag} d_{i\ell'\sigma'} d_{i\ell'\sigma}, 
\label{eq:H'}
\end{eqnarray}
where $i$ is on Ni sites, $\ell$ and $\ell'$ are 1 or 2. 
$U$ is the intra-orbital Coulomb repulsion, $U'$ is the inter-orbital Coulomb repulsion, 
$J$ is the Hund's rule coupling and $J'$ is the inter-orbital pair-hopping term. 
Throughout the present study, we take $U=8$ eV, $U'=6$ eV, and $J=J'=1$ eV. 

\begin{figure}
\begin{center}
\includegraphics[width=80mm]{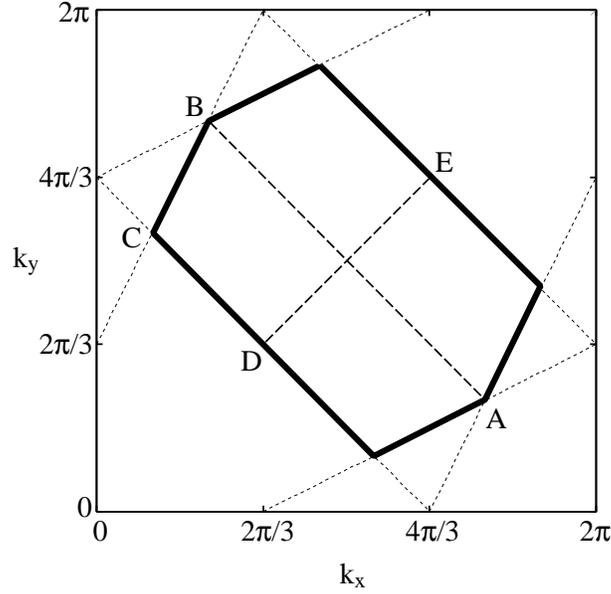}
\end{center}
\caption{
A typical reduced BZ for the diagonal stripe state is enclosed by the thick solid line. 
The wave vectors at some symmetry points are also indicated: 
A$(\frac{14}{9}\pi, \frac{4}{9}\pi)$, B$(\frac{4}{9}\pi , \frac{14}{9}\pi)$, 
C$(\frac{2}{9}\pi, \frac{10}{9}\pi)$, D$(\frac{2}{3}\pi, \frac{2}{3}\pi)$, 
E$(\frac{4}{3}\pi, \frac{4}{3}\pi)$.}
\label{Fig1}
\end{figure}

To deal with the many-body problem originating from $H'$, 
we adopt the HF approximation. 
A lot of HF calculations using Hubbard-type models
have been performed to study stripe ordering, 
so far~\cite{Poilblanc1989,Schulz1990,Kato1990,Zaanen1994,Zaanen1996,
Mizokawa1997,Ichioka1999,Kaneshita2001,Raczkowski2006,Kaneshita2008}. 
Here we introduce the mean fields 
\begin{equation}
\langle d_{i\ell\sigma}^{\dag} d_{i\ell\sigma} \rangle = \frac{1}{2}(n_{i\ell} + m_{i\ell} \sigma), 
\end{equation}
where the $n_{i\ell}$ and $m_{i\ell}$ are the electron number and the spin magnetic moment 
in orbital $\ell$ at Ni site $i$, respectively. 
We assume that the stripe ordering characterized 
by stripe vectors ${\mib Q}$'s is realized. 
In this case, the mean fields are expressed in the form: 
\begin{eqnarray}
n_{i\ell} &=& \sum_{\mib Q} e^{{\rm i} {\mib Q} \cdot {\mib r}_i} n_{{\mib Q}\ell}, \\ 
\label{eq:ni}
m_{i\ell} &=& \sum_{\mib Q} e^{{\rm i} {\mib Q} \cdot {\mib r}_i} m_{{\mib Q}\ell}, 
\label{eq:mi}
\end{eqnarray}
where $i$ is on Ni sites and $\ell$ is for Ni3$d$ orbitals, 
and ${\mib Q}$'s characterize the spatial periodicity of the stripe. 
In the present study, we restrict ourselves to diagonal stripe ordering 
expected from experimental results for La$_{5/3}$Sr$_{1/3}$NiO$_4$
and the above summation in ${\mib Q}$ 
is performed for ${\mib Q}=n \cdot {\mib Q}_s$ $(n=0, 1, 2)$, 
where ${\mib Q}_s=(\frac{2}{3}\pi, \frac{2}{3}\pi)$. 
$n_{{\mib Q}\ell} = n_{-{\mib Q}\ell}^*$ and 
$m_{{\mib Q}\ell} = m_{-{\mib Q}\ell}^*$ hold, 
since $n_{i\ell}$ and $m_{i\ell}$ are real quantities. 
As a result of the HF approximation, the total mean-field Hamiltonian 
$H_{\rm MF}=H_0 + H_{\rm MF}'$ is given in the form: 
\begin{equation}
H_{MF} = \sum_{{\mib k}} \sum_{\ell\ell'\sigma} \xi_{\ell\ell'}({\mib k}) 
c^{\dag}_{{\mib k}\ell\sigma} c_{{\mib k}\ell'\sigma} 
+ \sum_{{\mib k}} \sum_{\mib Q} \sum_{\ell\sigma}  
\Delta_{{\mib Q}\ell\sigma} d^{\dag}_{{\mib k}\ell\sigma} 
d_{{\mib k}-{\mib Q}\ell\sigma}  + E_0, 
\label{eq:H_MF}
\end{equation}
with 
\begin{eqnarray}
\Delta_{{\mib Q}\ell\sigma} &=& \frac{U}{2}(n_{{\mib Q}\ell} - m_{{\mib Q}\ell}\sigma) 
+ U'  \sum_{\ell' (\neq \ell)} n_{{\mib Q}\ell'} \nonumber \\
&& - \frac{J}{2} \sum_{\ell' (\neq \ell)} (n_{{\mib Q}\ell'} + m_{{\mib Q}\ell'} \sigma), \\
E_0 &=& -\frac{NU}{4} \sum_{\mib Q} \sum_{\ell} (|n_{{\mib Q}\ell}|^2-|m_{{\mib Q}\ell}|^2) 
-\frac{NU'}{2} \sum_{\mib Q} \sum_{\ell \neq \ell'} n_{{\mib Q}\ell} n_{{\mib Q}\ell'}^* \nonumber\\
&& + \frac{NJ}{4} \sum_{\mib Q} \sum_{\ell \neq \ell'} (n_{{\mib Q}\ell} n_{{\mib Q}\ell'}^* 
+ m_{{\mib Q}\ell} m_{{\mib Q}\ell'}^*). 
\label{eq:DeltaQ}
\end{eqnarray}
The first BZ is folded, since the spatial periodicity of the charge disproportionation 
is integer times of the original lattice periodicity. 
We express momentum ${\mib k}$ in the original BZ by using the stripe vector ${\mib Q}$ 
and reduced momentum ${\mib k}_0$ as ${\mib k}={\mib k}_0+{\mib Q}$, 
where ${\mib k}_0$ is restricted in the folded BZ. 
In the present study, we use the reduced BZ depicted in Fig.~\ref{Fig1}, 
to solve the HF equation for the stripe-ordered states 
with ${\mib Q}_s=(\frac{2}{3}\pi, \frac{2}{3}\pi)$. 
Thus the total mean-field Hamiltonian is expressed in the form: 
\begin{eqnarray}
H_{MF} &=&\sum_{{\mib k}_0} \sum_{{\mib Q}} 
\sum_{\ell\ell'\sigma} \xi_{\ell\ell'}({\mib k}_0+{\mib Q}) 
c^{\dag}_{{{\mib k}_0+\mib Q}\ell\sigma} c_{{\mib k}_0+{\mib Q}\ell'\sigma} \nonumber\\
&& + \sum_{{\mib k}_0} \sum_{{\mib Q},{\mib Q}'} \sum_{\ell\sigma} 
\Delta_{{\mib Q}-{\mib Q}'\ell\sigma} 
d^{\dag}_{{\mib k}_0+{\mib Q}\ell\sigma} d_{{\mib k}_0+{\mib Q}'\ell\sigma} + E_0. 
\label{eq:H_MF_2}
\end{eqnarray}
We should note that the summation in momentum ${\mib k}_0$ 
is restricted  only over the reduced BZ. 
By diagonalizing this mean-field Hamiltonian, we have the energy dispersions 
$E_{a, \sigma}({\mib k}_0)$ and unitary matrix $U_{{\mib Q}, \ell, a, \sigma}({\mib k}_0)$ 
for diagonalization, where $a$ is the index for the diagonalized bands. 
The self-consistency condition for mean fields is given by 
\begin{eqnarray}
n_{{\mib Q}\ell} &=& \frac{1}{N} \sum_{{\mib k}_0} \sum_{{\mib Q}'} \sum_{a,\sigma} 
U_{{\mib Q}', \ell, a, \sigma}^*({\mib k}_0) 
U_{{\mib Q}'+{\mib Q}, \ell, a, \sigma}({\mib k}_0) n_{a, \sigma}({\mib k}_0), \label{eq:nq}\\
m_{{\mib Q}\ell} &=& \frac{1}{N} \sum_{{\mib k}_0} \sum_{{\mib Q}'} \sum_{a,\sigma} 
\sigma U_{{\mib Q}', \ell, a, \sigma}^*({\mib k}_0) 
U_{{\mib Q}'+{\mib Q}, \ell, a, \sigma}({\mib k}_0) n_{a, \sigma}({\mib k}_0), \label{eq:mq}
\end{eqnarray}
where $n_{a, \sigma}({\mib k}_0)$ is the electron occupation number 
on band $a$ at momentum ${\mib k}_0$ with spin $\sigma$: 
$n_{a, \sigma}({\mib k}_0) = f(E_{a, \sigma}({\mib k}_0))$ 
($f(E)$ is the Fermi distribution function). 

Here we take account of the electron-lattice interaction, 
since it is natural to consider that lattice distortions occur cooperatively 
with the charge disproportionation accompanying the stripe ordering, 
as discussed in refs.~\ref{Zaanen1994}, \ref{Zaanen1996} and \ref{Kaneshita2008}. 
We denote the atom displacement at site $i$ by ${\mib u}({\mib r}_i)$. 
If we assume $t_{\ell, \ell'}({\mib r}) \propto |{\mib r}|^{-\alpha_{\ell\ell'}}$
($\alpha_{\ell\ell'}=\alpha_{\ell'\ell}$), 
the change of the hopping parameters due to the atom displacement 
is evaluated approximately 
by $\delta t_{\ell,\ell'}({\mib r}_{ij}) = - \alpha_{\ell\ell'} t_{\ell,\ell'}({\mib r}_{ij})
\frac{{\mib r}_{ij} \cdot {\mib u}_{ij}}{|{\mib r}_{ij}|^2}$, 
where ${\mib u}_{ij} = {\mib u}({\mib r}_i) - {\mib u}({\mib r}_j) $. 
Hereafter, for ${\mib u}_{ij}$ and ${\mib r}_{ij}$, 
we use units normalized by the lattice constant. 
The modification to the Hamiltonian $H_0$ due to the lattice distortions is given by 
\begin{equation}
H_{\rm l.d.} = \sum_{i,j} \sum_{\ell\ell'\sigma} \Bigl[ - \alpha_{\ell\ell'} t_{\ell,\ell'}({\mib r}_{ij}) 
\frac{{\mib r}_{ij} \cdot {\mib u}_{ij}}{|{\mib r}_{ij}|^2} \Bigr] 
c^{\dag}_{i\ell\sigma} c_{j\ell'\sigma}.  
\label{eq:H_ld}
\end{equation}
Throughout  the present study, we take 
$\alpha_{\ell\ell'} \equiv \alpha_{dp}=3.5$ 
for $(\ell, \ell')=(1$ or $2, 3$ or $4)$, $(3$ or $4, 1$ or $2)$ (Ni-O bonds), 
and $\alpha_{\ell\ell'} \equiv \alpha_{pp}=2$ 
for $(\ell, \ell')=(3$ or $4, 3$ or $4)$ (O-O bonds)~\cite{Harrison1989}. 
We use the following form for the lattice elastic energy, 
\begin{equation}
E_{\rm L}[{\mib u}_{ij}] = \frac{1}{2} \sum_{(i,j)} K_{A_i,A_j}(|{\mib r}_{ij}|) |{\mib u}_{ij}|^2,  
\label{eq:E_L}
\end{equation}
where $A_i$ and $A_j$ denote the atoms (Ni or O) at sites $i$ and $j$, respectively, 
and $K_{A_i,A_j}(|{\mib r}_{ij}|)$ is the elastic constant 
of the bond between atoms $A_i$ and $A_j$. 
The summation with $(i, j)$ means that with respect to all pairs of sites. 
Minimizing the total energy 
\begin{equation}
E_{\rm tot} = \langle H_0 + H'+ H_{\rm l.d.} + E_{\rm L} \rangle
\label{eq:E_tot}
\end{equation}
with respect to the lattice distortion ${\mib u}_{ij}$ 
and using the Hellmann-Feynman theorem, 
we have the following equation for determining ${\mib u}_{ij}$: 
\begin{equation}
{\mib u}_{ij} = \frac{1}{K_{A_i,A_j}(|{\mib r_{ij}}|)} 
\sum_{\ell\ell'\sigma} \alpha_{\ell\ell'} t_{\ell,\ell'}({\mib r}_{ij})  
\frac{{\mib r}_{ij}}{|{\mib r}_{ij}|^2} 
\langle c_{i\ell\sigma}^{\dag} c_{j\ell'\sigma} 
+ c_{j\ell'\sigma}^{\dag} c_{i\ell\sigma} \rangle, \label{eq:uij}
\end{equation}
where the summation in $\ell$ ($\ell'$) is restricted to the orbitals 
on atom $A_i$ ($A_j$) at site $i$ ($j$). 
As can be seen easily, this condition eq.~($\ref{eq:uij}$) is not effective 
for pairs of sites $i$ and $j$ for which $t_{\ell,\ell'}({\mib r}_{ij})$ equals zero 
for any choice of $\ell$ and $\ell'$. 
In the present study, since we take only hoppings 
for nearest-neighbor Ni$3d$-O$2p$ and O$2p$-O$2p$ bonds, 
only ${\mib u}_{ij}$'s for these bonds are determined by eq.~(\ref{eq:uij}), 
while ${\mib u}_{ij}$'s for the other bonds are not. 

Here we assume that the lattice distortions possess the same spatial periodicity 
as the stripe order, and then can expand ${\mib u}_{ij}$ in the form 
\begin{equation}
{\mib u}_{ij} = \sum_{{\mib Q}, {\mib Q}'} 
{\mib u}_{{\mib Q}\ell,{\mib Q}'\ell'}({\mib r}_{ij}) 
e^{{\rm i} ({\mib Q} \cdot {\mib r}_i - {\mib Q}' \cdot {\mib r}_j)}, 
\label{eq:uij_2}
\end{equation}
where $\ell$ and $\ell'$ denote the atomic orbitals at sites $i$ and $j$, respectively. 
$H_{\rm l.d.}$ is expressed in the momentum representation as 
\begin{equation}
H_{\rm l.d.} = \sum_{{\mib k}_0} \sum_{{\mib Q},{\mib Q}'} 
\sum_{\ell\ell'\sigma} \eta_{\ell\ell'}({\mib k}_0; {\mib Q}, {\mib Q}') 
c_{{\mib k}_0+{\mib Q}\ell\sigma}^\dag c_{{\mib k}_0+{\mib Q}'\ell'\sigma},  
\label{eq:H_ld_2}
\end{equation}
with 
\begin{equation}
\eta_{\ell\ell'}({\mib k}_0; {\mib Q},{\mib Q}') = - \alpha_{\ell\ell'} \sum_{\mib r} 
t_{\ell, \ell'}({\mib r}) \frac{e^{-{\rm i} {\mib k}_0 \cdot {\mib r}}}{|{\mib r}|^2}  
\sum_{{\mib Q}''} [{\mib r} \cdot {\mib u}_{{{\mib Q}-{\mib Q}''}\ell, {{\mib Q}'-{\mib Q}''}\ell'}({\mib r})]
e^{- {\rm i} {\mib Q}'' \cdot {\mib r}}. 
\label{eq:eta}
\end{equation}
The energy minimization condition eq.~(\ref{eq:uij}) is given 
in the momentum representation by 
\begin{eqnarray}
{\mib u}_{{\mib Q}\ell,{\mib Q}'\ell'}({\mib r}_{ij}) = \frac{1}{K_{A_i,A_j}(|{\mib r}_{ij}|)} 
\sum_{mm'} \alpha_{mm'} t_{m,m'}({\mib r}_{ij}) \frac{{\mib r}_{ij}}{|{\mib r}_{ij}|^2} \nonumber \\
\times \frac{1}{N} \sum_{{\mib k}_0} \sum_{a\sigma} [
e^{-{\rm i} {\mib k}_0 \cdot {\mib r}_{ij}} U_{-{\mib Q},m,a,\sigma}^*({\mib k}_0) 
U_{-{\mib Q}',m',a,\sigma}({\mib k}_0) \nonumber\\
+ e^{{\rm i} {\mib k}_0 \cdot {\mib r}_{ij}} U_{{\mib Q}',m',a,\sigma}^*({\mib k}_0) 
U_{{\mib Q},m,a,\sigma}({\mib k}_0)] n_{a, \sigma}({\mib k}_0), 
\label{eq:uqq}
\end{eqnarray}
where $A_i$ and $A_j$ denote atoms on which orbitals $\ell$ and $\ell'$ are, respectively, 
and the summation in $m$ and $m'$ is restricted to the orbitals on the atoms $A_i$ and $A_j$, respectively. 
$U_{{\mib Q},\ell,a,\sigma}({\mib k}_0)$ is the diagonalization matrix for $H_{\rm MF} + H_{\rm l.d.}$. 
As easily seen, ${\mib u}_{{\mib Q}\ell,{\mib Q}'\ell'}({\mib r}_{ij})$ satisfies the following equalities, 
\begin{eqnarray}
{\mib u}_{{\mib Q}\ell,{\mib Q}'\ell'}({\mib r}_{ij}) 
&=& {\mib u}_{-{\mib Q}\ell,-{\mib Q}'\ell'}^*({\mib r}_{ij}) \nonumber \\ 
&=& - {\mib u}_{-{\mib Q}'\ell',-{\mib Q}\ell}(-{\mib r}_{ij}). 
\end{eqnarray}

If we do not assume any lattice distortions, 
self-consistent solutions are determined by diagonalizing $H_{\rm MF}$ 
in eq.~(\ref{eq:H_MF_2}) and using the self-consistent conditions 
eqs.~(\ref{eq:nq}) and (\ref{eq:mq}). 
If we assume lattice distortions, self-consistent solutions are determined 
by diagonalizing $H_{\rm MF} + H_{\rm l.d.}$ from eqs.~(\ref{eq:H_MF_2}) 
and (\ref{eq:H_ld_2}) and using the self-consistency conditions 
eqs.~(\ref{eq:nq}), (\ref{eq:mq}) and (\ref{eq:uqq}). 

For the undoped case, we expect the checkerboard-type antiferromagnetic ground state. 
Such a ground state is obtained within the same formulation 
by taking ${\mib Q}_s=(\pi, \pi)$ and ${\mib Q} = n \cdot {\mib Q}_s$ $(n=0, 1)$. 
In this case, the reduced BZ is a square whose corners 
are $(\pm \pi, 0)$ and $(0, \pm \pi)$, in the wave vector space. 

We should note that the identity 
\begin{equation}
{\mib u}_{ij}+{\mib u}_{jk}+{\mib u}_{ki}=0
\label{eq:uijujk}
\end{equation}
should always hold for any choice of lattice sites $i$, $j$ and $k$, 
as easily shown from the definition ${\mib u}_{ij} = {\mib u}({\mib r}_i) - {\mib u}({\mib r}_j)$. 
However, this identity and the energy minimization condition 
eq.~(\ref{eq:uij}) are not necessarily consistent with each other in general. 
Rigorously speaking, we should minimize the total energy eq.~(\ref{eq:E_tot}) 
under the condition eq.~(\ref{eq:uijujk}), but this is a rather complex and difficult task. 
In \S~\ref{Sc:Stripe}, we present a possible resolution for this difficulty. 

\subsection{RIXS formula}
\label{Sc:Formula}

To calculate the RIXS intensity, we use the useful analytic formula previously presented 
by Nomura and Igarashi~\cite{Nomura2004, Nomura2005, Igarashi2006}. 
This analytic formula has been applied to insulating copper 
oxides~\cite{Nomura2004, Nomura2005, Igarashi2006,Takahashi2008}, 
NiO~\cite{Takahashi2007}, LaMnO$_3$~\cite{Semba2008}, La$_2$NiO$_4$~\cite{Takahashi2009}, 
using realistic electronic structures, and explained the shape and momentum-transfer 
dependence of RIXS charge excitation spectra semiquantitatively. 
Here we outline the derivation of their formula. 

The Hamiltonian for the interaction between x-ray and electrons is given by 
\begin{eqnarray}
H_x &=& \sum_{{\mib q}, {\mib e}} \tilde{H}_x ({\mib q},{\mib e}) \alpha_{{\mib q}{\mib e}} + {\rm H. c.}, \\
\tilde{H}_x({\mib q},{\mib e}) &=& \sum_{{\mib k}, \sigma} w({\mib q}, {\mib e}) 
p'{}^{\dag}_{{\mib k}+{\mib q}\sigma} s_{{\mib k}\sigma}. 
\end{eqnarray}
where $s_{{\bf k}\sigma}$ and $p'_{{\bf k}\sigma}$ are the annihilation operators 
for the transition-metal $1s$ and $4p$ electrons with momentum ${\mib k}$ and spin $\sigma$, 
$\alpha_{{\mib q}{\mib e}}$ is the annihilation operator for x-ray photons 
with wave vector ${\mib q}$ and polarization vector ${\mib e}$. 
The matrix element $w({\mib q}, {\mib e})$ is given by
\begin{equation}
w({\mib q}, {\mib e})=-\frac{e}{m} \sqrt{\frac{2 \pi}{|{\mib q}|}}
\langle 4p|e^{{\rm i}{\mib q}\cdot{\mib r}}{\mib e}\cdot{\mib p}|1s \rangle, 
\end{equation}
in units of $c=\hbar=1$ 
($c$: light velocity, $\hbar$: Planck constant divided by $2\pi$), 
where $e$ and $m$ are charge and mass of the electron, respectively. 

Following Nozi\`eres and Abrahams~\cite{Nozieres1974}, 
we calculate the inelastic scattering intensity. 
We denote the initial ground state of the electronic system 
without any photon by $|0 \rangle$ in the infinite past time $t=-\infty$. 
We assume that this state $|0 \rangle$ absorbs an incident photon 
(momentum ${\mib q}_i$, energy $\omega_i$ and polarization ${\mib e}_i$), 
and emits a photon (momentum ${\mib q}_f$, energy $\omega_f$ 
and polarization ${\mib e}_f$), before a time $t_0$, 
and the electronic system is in an excited state $|\psi(t_0) \rangle $ at $t_0$. 
The amplitude for the state $|\psi(t_0) \rangle $ is calculated 
within the second order perturbation theory in $H_x$: 
\begin{equation}
| \psi(t_0) \rangle = - \int_{-\infty}^{t_0} du \int_{-\infty}^u dt \, K(t_0, u) H_x 
K(u, t) H_x K(t, -\infty) \alpha_{{\mib q}_i{\mib e}_i}^{\dag} |0 \rangle, 
\end{equation}
where $K(u, t)$ is the time evolution operator in the case of $H_x=0$, 
and $t$ and $u$ are the times of the photon absorption and emission, respectively. 
The total number of x-ray photons generated before $t_0$ 
with wave vector ${\mib q}_f$ and polarization ${\mib e}_f$ is 
$N_f(t_0)= \langle \psi(t_0)|\alpha^{\dag}_{{\mib q}_f{\mib e}_f} 
\alpha_{{\mib q}_f{\mib e}_f}|\psi(t_0) \rangle$. 
Inelastic x-ray scattering spectra are regarded as the number of photons 
generated in a unit time. Thus, we obtain the scattering intensity by deriving 
$N_f(t_0)$ with respect to $t_0$ and contracting photon annihilation and creation operators: 
\begin{equation}
W(q_i{\mib e}_i; q_f{\mib e}_f) = 
\int_{-\infty}^{\infty}{du'}\,\int_{-\infty}^{u'} {dt'}\, \int_{-\infty}^0{dt} 
e^{{\rm i}\omega_i(t'-t)}e^{-{\rm i}\omega_fu'} S(t, 0; u', t'), 
\label{eq:rixssp1}
\end{equation}
with 
\begin{equation}
S(t, u; u', t') =
\langle 0|\tilde{H}_x^{\dag}(t'; {\mib q}_i,{\mib e}_i)\tilde{H}_x(u'; {\mib q}_f,{\mib e}_f) 
\tilde{H}_x^{\dag}(u; {\mib q}_f,{\mib e}_f)\tilde{H}_x(t; {\mib q}_i,{\mib e}_i) |0 \rangle, 
\label{eq:intensity}
\end{equation}
where $\tilde{H}_x(t; {\mib q},{\mib e})$ is the interaction representation 
of $\tilde{H}_x({\mib q},{\mib e})$. 
The function $S(t, u; u', t')$ can be calculated by the Keldysh formalism. 

For the case of RIXS at the transition-metal $K$ edges, 
we introduce the following approximations: 
(i) We take completely flat dispersion for the $1s$ band, 
since the $1s$ electrons are strongly localized in the inner $K$ shell. 
(ii) We use a free-electron model for the 4$p$ electrons, 
since the transition-metal 4$p$ orbitals are expected to much extend in space. 
In the present study, we take simply a cosine-shaped band for the $4p$ electrons. 
This is justified by the fact that the excited $4p$ electron plays only a role 
of ``spectator"~\cite{Kim2007}, as far as we discuss the momentum dependence of RIXS spectra. 
Of course, detailed $4p$ electron energy dispersions are necessary for more quantitatively precise 
discussions on $K$ edge absorption spectra and dependences on the incident photon energy 
and polarization.
(iii) Since transition-metal $d$ electrons also possess a localized nature, 
the 1$s$ core-hole potential (whose absolute value equals the Coulomb integral 
$V_{1s-d}$ between the $1s$ and $d$ electrons) is expected to be rather strong. 
Nevertheless, we use the Born approximation for $V_{1s-d}$. 
The Born approximation was partly justified by taking account 
of multiple-scattering processes for the case of La$_2$CuO$_4$~\cite{Igarashi2006}. 
On the basis of these assumptions, we can perform the integrals with respect 
to time variables in eq.~(\ref{eq:rixssp1}), and have
\begin{equation}
W(q_i{\mib e}_i; q_f{\mib e}_f) = 
4\biggl| \frac{V_{1s-d}}{N} \sum_{\mib k} 
\frac{w({\mib q}_i, {\mib e}_i) w({\mib q}_f, {\mib e}_f)^*}
{\gamma(\omega_i; {\mib k}) \gamma(\omega_f; {\mib k})}\biggr|^2 
Y_d({\mib q}_i-{\mib q}_f, \omega_i-\omega_f), 
\label{eq:rixssp2}
\end{equation}
with
\begin{eqnarray}
\gamma(\omega; {\mib k}) &=& \omega+\varepsilon_{1s}+{\rm i}\Gamma_{1s}-\varepsilon_{4p}({\mib k}), 
\end{eqnarray}
where $\varepsilon_{\rm 1s}$, $\Gamma_{\rm 1s}$ and $\varepsilon_{4p}({\mib k}) $ 
are the $1s$ energy level, the decay rate of  the $1s$ core hole, and the $4p$ band energy, respectively. 
$Y_d({\mib q}, \omega)$ is the Fourier transform of the dynamical charge-density correlation 
function for $d$ electrons: 
\begin{equation}
Y_d({\mib q}, t)= \langle \rho_{d\,\mib q}(t) \rho_{d\,-\mib q}(0) \rangle, 
\end{equation}
where $\rho_{d\mib q}(t)$ is the Heisenberg representation 
of the following electron density operator for $d$ electrons, 
\begin{equation}
\rho_{d\mib q} = \sum_{{\mib k}, \sigma} d_{{\mib k}\sigma}^{\dag} d_{{\mib k}+{\mib q}\sigma}. 
\end{equation}
We should note that the $1s$ core hole plays a role of a localized testing 
charge perturbing the $d$ electrons through the $1s$-$d$ Coulomb interaction. 
As easily seen from the above discussions, applying the Born approximation 
to $V_{1s{\rm -}d}$ is equivalent to taking account only of the linear response 
of $d$ electrons to the perturbation due to the $1s$ core hole charge. 
The fluctuation dissipation theorem relates $Y_d({\mib q}, \omega)$
to the charge susceptibility of $d$ electrons $\chi_d({\mib q}, \omega)$: 
\begin{equation}
Y_d({\mib q}, \omega)= 2 (1-e^{-\omega/T})^{-1} {\rm Im} \chi_d({\mib q}, \omega+{\rm i}0). 
\end{equation}

In the present study, the incident photon energy $\omega_i$ is set near the $K$ absorption 
energy, i.e., $\omega_i \approx \varepsilon_{4p}({\mib k}_{\rm edge}) - \varepsilon_{1s} $. 
Then $\gamma(\omega_i; {\mib k}) $ becomes small and the RIXS intensity 
$W(q_i{\mib e}_i; q_f{\mib e}_f)$ is resonantly enhanced. 
The resolution of the RIXS spectra as a function 
of photon energy loss $\omega= \omega_i -\omega_f$ is determined 
by that of $Y_d({\mib q}, \omega)$, i.e., 
the decay rate of charge excitations of $d$ electrons, 
rather than by the core hole decay rate $\Gamma_{1s}$. 

To proceed further, we have to evaluate explicitly $Y_d({\mib q}, \omega)$. 
Specifically we take the HF approximation to describe the stripe-ordered ground state 
of the $d$ electrons, and the RPA to take account of the electron correlations in the intermediate state. 
RPA means that we neglect any couplings among various modes corresponding 
to different wave vectors ${\mib q}$'s, in other words, we assume various excitation 
modes with different ${\mib q}$'s are separately renormalized by electron correlations. 
Consequently we have the final expression for stripe-ordered states: 
\begin{eqnarray}
W(q_i{\mib e}_i, q_f{\mib e}_f) &=& 8 \pi \biggl| \frac{V_{1s-d}}{N} \sum_{\mib k} 
\frac{w({\mib q}_i, \alpha_i)w^*({\mib q}_f, \alpha_f)}
{\gamma(\omega_i; {\mib k}) \gamma(\omega_f; {\mib k})} \biggr|^2 \nonumber\\
&&\times \frac{1}{N} \sum_{{\mib k}_0} \sum_{aa'\sigma} 
\delta( \omega + E_{a,\sigma}({\mib k}_0) - E_{a',\sigma}(\{{\mib k}_0+{\mib q}\})) 
n_{a, \sigma}({\mib k}_0) [1 - n_{a', \sigma}(\{{\mib k}_0+{\mib q}\})] \nonumber\\
&&\times \biggl| \sum_{\ell\ell'} \sum_{{\mib Q},{\mib Q}'} 
\Lambda_{\ell'\ell}^{\sigma}({\mib Q}+\tilde{\mib Q}_{{\mib k}_0+{\mib q}}-{\mib Q}') 
U_{{\mib Q},\ell,a,\sigma}({\mib k}_0) U_{{\mib Q}',\ell',a',\sigma}^*(\{{\mib k}_0+{\mib q}\}) \biggr|^2,  
\end{eqnarray}
where $\tilde{\mib Q}_{{\mib k}_0+{\mib q}}$ is the stripe vector 
by which the wave vector ${\mib k}_0+{\mib q}$ is pulled back into the reduced BZ, 
and  then the wave vector ${\mib k}_0+{\mib q}$ is reduced to the wave vector 
$\{{\mib k}_0+{\mib q}\}$ in the reduced BZ, i.e., 
$\{{\mib k}_0+{\mib q}\} \equiv  {\mib k}_0+{\mib q} - \tilde{\mib Q}_{{\mib k}_0+{\mib q}}$. 
$\Lambda_{\ell'\ell}^{\sigma}({\mib Q}) $ is the vertex function including 
electron correlation effects, which we calculate within RPA 
with respect to the Coulomb interaction $H'$ of eq.~(\ref{eq:H'}). 
On the other hand, if we turn off the RPA corrections 
by setting $\Lambda_{\ell'\ell}^{\sigma}({\mib Q}) = \delta_{\mib Q} \delta_{\ell'\ell}$, 
then we can extract simple band-to-band charge excitations. 

Numerical diagonalization technique using finite cluster models is another promising method 
for calculating RIXS intensity~\cite{Tsutsui1999,Tsutsui2003,Okada2006}, 
in the sense that it can take full account 
of electron correlations without any approximations, 
but may have severe difficulty in analyzing detailed momentum dependences 
for such stripe-ordered systems as La$_{5/3}$Sr$_{1/3}$NiO$_4$, 
since the unit cell in stripe states contains a relatively large number of atoms. 

\section{Numerical Results}
\label{Sc:Numerical} 

\subsection{
Case of undoped antiferromagnetic insulating state: $x=0$, ${\mib Q}=(\pi, \pi)$}
\label{Sc:AF}

\begin{figure}
\begin{center}
\includegraphics[width=90mm]{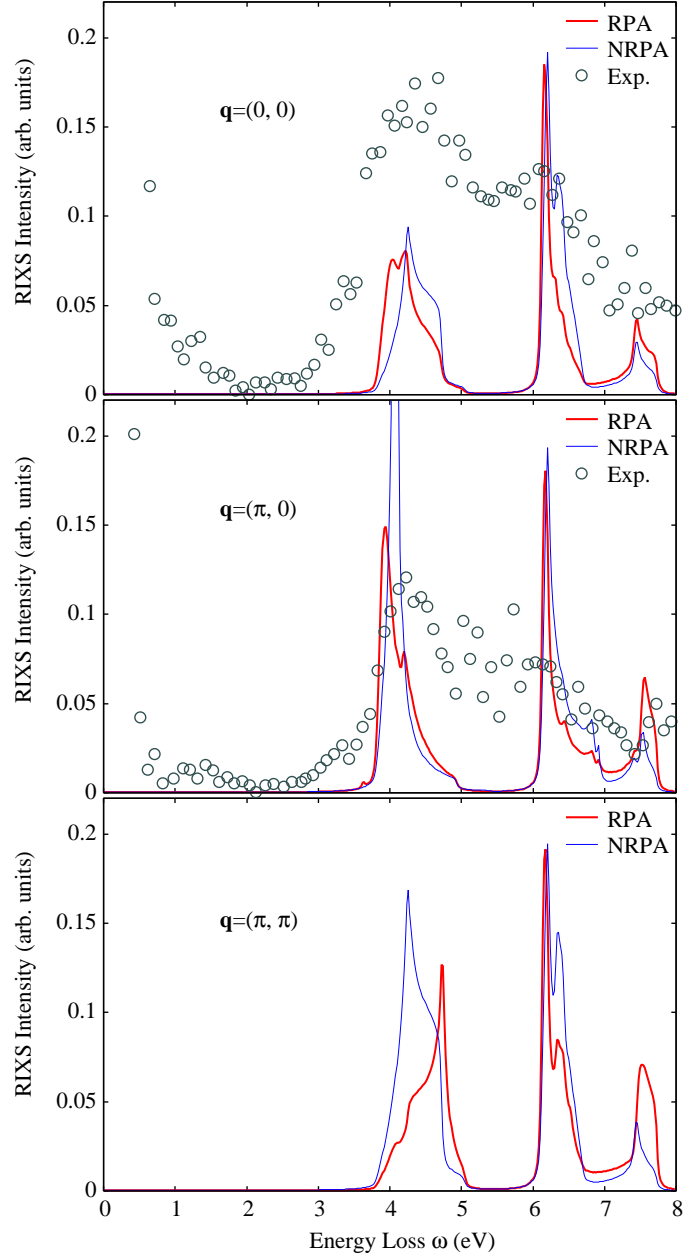}
\end{center}
\caption{
(Color online)
Calculated results of RIXS spectra for three momentum transfers 
${\mib q}=(0, 0), (\pi, 0), (\pi, \pi)$, and comparison with experimental results 
at ${\mib q}=(0, 0), (\pi, 0)$. 
The empty circles represent experimental data read from Ref.~\ref{Collart2006}. 
The thick red and thin blue lines represent the results calculated 
with and without RPA, respectively. 
}
\label{Fig2}
\end{figure}

Before proceeding to the cases for stripe ordered states, 
we present the results for the undoped antiferromagnetic state. 
Based on the formulation in \S~\ref{Sc:Formulation}, 
the antiferromagnetic ground state is obtained: 
the electron occupation number and total staggered spin moment 
are $n_i = 2.24$ and $m_i = 1.74$ (in units of $\mu_{\rm B}$) 
at each Ni site $i$, respectively, where we have not assumed lattice distortions. 
The calculated RIXS spectra for three momentum transfers ${\mib q}=(0, 0), (\pi, 0)$ 
and $(\pi, \pi)$ are presented in Fig.~\ref{Fig2}. 
We find three spectral features around the energy loss $\omega=4$, $6$ and $7.5$ eV. 
Overall, the spectral peak positions in the calculation seem consistent 
with the experimental results semiquantitatively. 
On the other hand, the total integrated intensity at ${\mib q}=(0, 0)$ 
seems clearly larger than that at ${\mib q}=(\pi, 0)$ in the experiment, 
while it does not seem in the theory. 
This inconsistency may be resolved by taking account of multiple scatterings 
beyond the Born approximation for the core-hole potential, 
since the low-energy intensity at ${\mib q}=(0, 0)$ seems more enhanced 
due to the multiple scatterings than at ${\mib q}=(\pi, 0)$, 
according to the study for La$_2$CuO$_4$~\cite{Igarashi2006}. 
In addition, for more complete quantitative consistency in spectral shape, 
we should use not such simple tight-binding electronic structures 
but more precise electronic band structures 
for all of the Ni3$d$, O2$p$ and Ni4$p$ states. 
Comparing the peak positions between the cases with and without RPA, 
the 4 eV peak is shifted to the high energy region due to electron correlations 
at ${\mib q}=(\pi, \pi)$, while it is not at ${\mib q}=(0, 0)$ and $(\pi, 0)$. 

\begin{figure}
\begin{center}
\includegraphics[width=90mm]{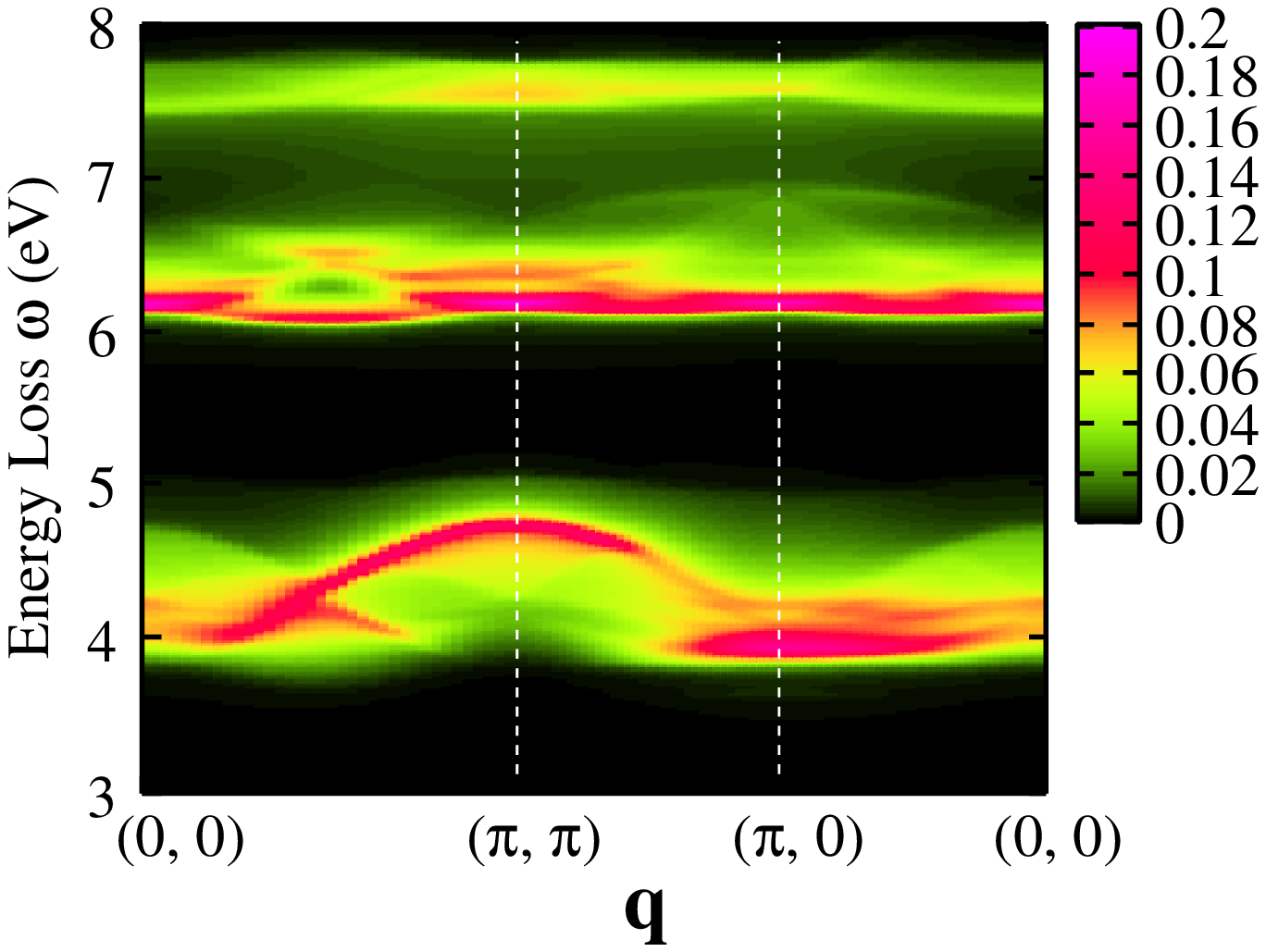}
\end{center}
\caption{
(Color online)
Intensity plot of the RIXS spectra calculated for the undoped antiferromagnetic state. 
The horizontal axis represents the momentum transfer 
along the symmetry lines of the square BZ, 
and the vertical axis represents the energy loss of photons.}
\label{Fig3}
\end{figure}

Here, we present more detailed momentum dependence 
of the RIXS spectra than presented in the previous work 
by Takahashi et al.~\cite{Takahashi2009}. 
The detailed momentum dependence of RIXS spectra along the symmetry lines 
in the first BZ is shown by the intensity plot in Fig.~\ref{Fig3}. 
Our calculation suggests the possibility that the 4 eV peak 
exhibits stronger dispersion along the line $(0, 0)$-$(\pi, \pi)$
rather than along the line $(0, 0)$-$(\pi, 0)$. 
This contrasts strongly with the case of La$_2$CuO$_4$, 
in which the 2 eV peak shows strong dispersion along $(0, 0)$-$(\pi, 0)$~\cite{Kim2002}. 

\begin{figure}
\begin{center}
\includegraphics[width=90mm]{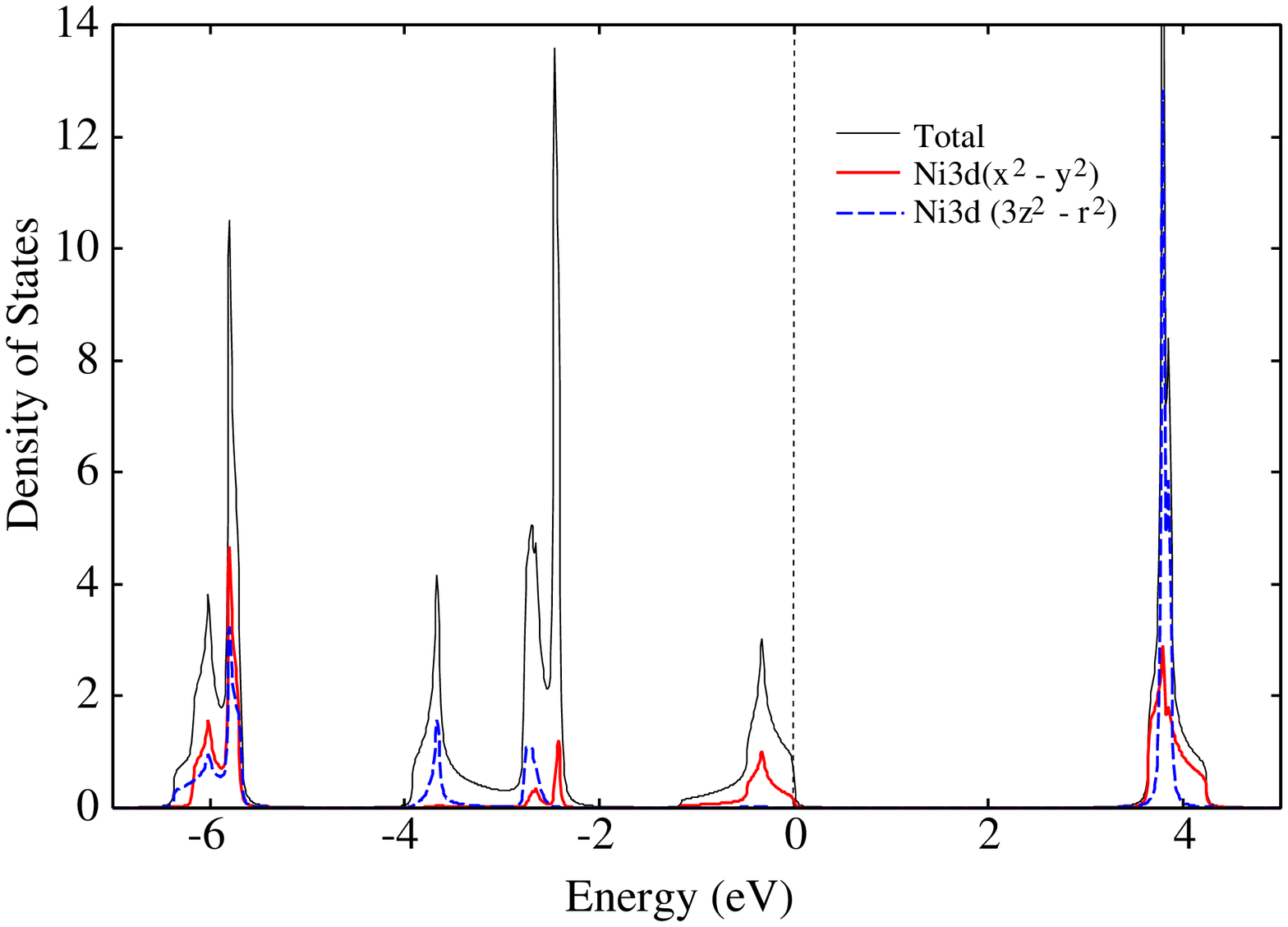}
\end{center}
\caption{
(Color online)
Total and Ni$3d$ partial density of states are depicted. 
The total DOS, and the partial DOS for Ni$3d_{x^2-y^2}$ and Ni$3d_{3z^2-r^2}$ 
orbitals are represented by thin black, thick red solid and thick blue broken lines, 
respectively. }
\label{Fig4}
\end{figure}

To see the microscopic origin of the RIXS weights, 
we present the results of density of states (DOS) in Fig.~\ref{Fig4}. 
The insulating gap energy, which corresponds to the charge transfer energy 
between the Ni$3d$ and O$2p$ bands, is about 4 eV, 
being consistent with photoemission experiments~\cite{Eisaki1992}. 
The 4 eV RIXS spectral weight corresponds to the charge transfer excitations 
from the Ni$3d$-O$2p$ anti-bonding band to the Ni$3d$ upper Hubbard band. 
The 6 eV peak corresponds to the charge transfer excitation 
from the Ni-O bonding band to the upper Hubbard band. 

\subsection{
Case of doped stripe-ordered states: 
$x=\frac{1}{3}$, ${\mib Q}_s=(\frac{2}{3}\pi, \frac{2}{3}\pi)$}
\label{Sc:Stripe}

As mentioned in the last paragraph of \S~\ref{Sc:Model}, 
it may be impossible to determine all ${\mib u}_{ij}$'s 
by solving eq.~(\ref{eq:uij}) consistently with eq.~(\ref{eq:uijujk}). 
For example, let us focus on a Ni atom and its two neighboring O atoms 
whose two Ni-O bonds cross perpendicularly each other. 
If those two Ni-O bonds shrink (the O-Ni-O angle remains perpendicular), 
then the O-O bond length determined from eq.~(\ref{eq:uijujk}) necessarily shrinks. 
However, this shrinkage of the O-O bond is not necessarily 
consistent with ${\mib u}_{ij}$ determined for that O-O bond 
by using the condition eq.~(\ref{eq:uij}). 
For this difficulty, we may take the following way: 
firstly we determine ${\mib u}_{ij}$ for each nearest-neighbor Ni-O bond 
by using the condition eq.~(\ref{eq:uij}) (or equivalently eq.~(\ref{eq:uqq})), 
and then we determine ${\mib u}_{ij}$ for each nearest-neighbor O-O bond 
by using eq.~(\ref{eq:uijujk}). 
Fortunately, this treatment seems valid, as inferred from the following discussions. 
As far as we studied, if we determine all ${\mib u}_{ij}$'s (including those for O-O bonds) 
from eq.~(\ref{eq:uij}) (or equivalently eq.~(\ref{eq:uqq})) without using the condition eq.~(\ref{eq:uijujk}), 
then the electronic structure and ${\mib u}_{ij}$'s only negligibly depend on $K_{\rm O,O}$, 
while they depend strongly on $K_{\rm Ni,O}$. 
This means that the lattice distortions are strongly dominated by changes of Ni-O bond length 
and the contributions from O-O bonds are almost energetically negligible. 
Therefore, to determine ${\mib u}_{ij}$ for each nearest-neighbor O-O bond, 
we may neglect the condition eq.~(\ref{eq:uij}) for O-O bonds, 
and instead of that, we should use eq.~(\ref{eq:uijujk}). 
The condition eq.~(\ref{eq:uijujk}) for O-O bonds reduces in the momentum representation to 
\begin{eqnarray}
{\mib u}_{{\mib Q}3,{\mib Q}'4} \Bigl( \pm \frac{\hat{x}}{2} \pm \frac{\hat{y}}{2} \Bigr)
= \sum_{{\mib Q}''} \Bigl[ {\mib u}_{{\mib Q}+{\mib Q}''3,{\mib Q}''\ell} \Bigl( \pm \frac{\hat{x}}{2} \Bigr) 
e^{ {\rm i} {\mib Q}'' \cdot (\pm \frac{\hat{x}}{2}) } \delta_{{\mib Q}'} 
+ {\mib u}_{{\mib Q}''\ell,{\mib Q}'+{\mib Q}''4} \Bigl( \pm \frac{\hat{y}}{2} \Bigr) 
e^{ {\rm i} {\mib Q}'' \cdot (\pm \frac{\hat{y}}{2}) } \delta_{\mib Q} \Bigr], 
\label{eq:upp1}
\\
{\mib u}_{{\mib Q}4,{\mib Q}'3} \Bigl( \pm \frac{\hat{x}}{2} \pm \frac{\hat{y}}{2} \Bigr)
= \sum_{{\mib Q}''} \Bigl[ {\mib u}_{{\mib Q}+{\mib Q}''4,{\mib Q}''\ell} \Bigl( \pm \frac{\hat{y}}{2} \Bigr) 
e^{ {\rm i} {\mib Q}'' \cdot (\pm \frac{\hat{y}}{2}) } \delta_{{\mib Q}'} 
+ {\mib u}_{{\mib Q}''\ell,{\mib Q}'+{\mib Q}''3} \Bigl( \pm \frac{\hat{x}}{2} \Bigr) 
e^{ {\rm i} {\mib Q}'' \cdot (\pm \frac{\hat{x}}{2}) } \delta_{\mib Q} \Bigr], 
\label{eq:upp2}
\end{eqnarray} 
where $\ell$ is Ni $d$ orbital (i.e., $\ell= 1$ or $2$), 
and the double signs in front of $\frac{\hat{x}}{2}$ (or $\frac{\hat{y}}{2}$) 
correspond in each equation. 

\begin{figure}
(a)
\begin{center}
\includegraphics*[width=70mm]{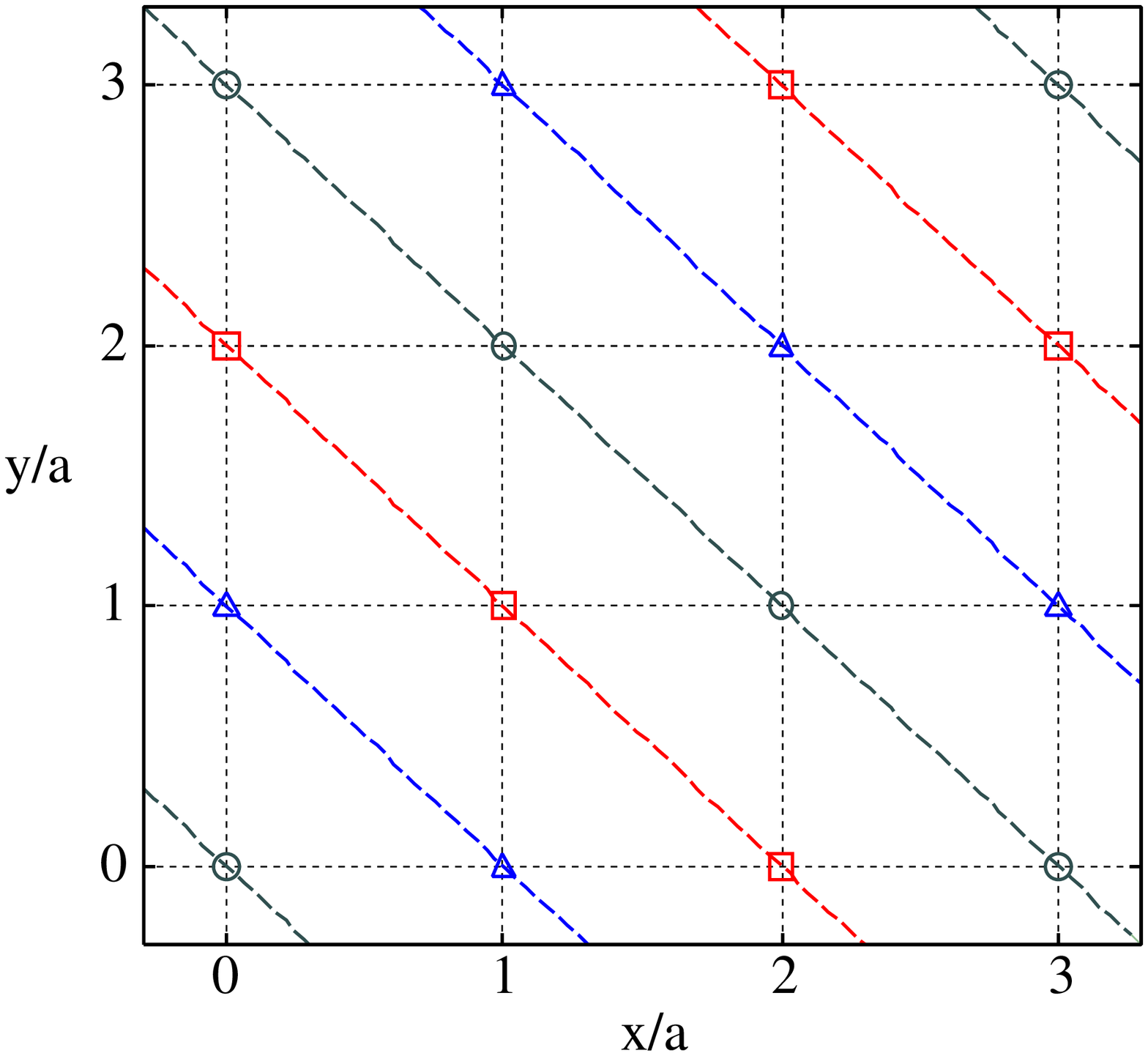}
\end{center}
(b) \begin{tabular}{|l|rrr|} \hline
HF stripe solution & $n_{\bigcirc}$ & $n_{\triangle}$ & $n_{\square}$ \\
& $m_{\bigcirc}$ & $m_{\triangle}$ & $m_{\square}$ \\ \hline\hline
(I): no lattice distortions & 2.16 & 2.20 & 2.16 \\
& 1.59 & -1.77 & 1.59 \\ \hline
(II): $K_{\rm Ni,O} = 30$ eV, & 2.26 & 2.37 & 2.34 \\
\hspace{5mm} not using eqs.~(\ref{eq:upp1}) and (\ref{eq:upp2}) & 0.98 & -1.45 & 1.40 \\ \hline
(III):  $K_{\rm Ni,O}= 30$ eV, & 2.30& 2.39 & 2.35 \\ 
\hspace{5mm} using eqs.~(\ref{eq:upp1}) and (\ref{eq:upp2}) & 1.06 & -1.42 & 1.31 \\ \hline 
(IV): $K_{\rm Ni,O}=60$ eV, & 2.22 & 2.29 & 2.27 \\ 
\hspace{5mm} using eqs.~(\ref{eq:upp1}) and (\ref{eq:upp2}) & 1.24 & -1.64 & 1.50 \\ \hline
\end{tabular}
\caption{
(Color online)
(a) Schematic figure of diagonal stripe states with ${\mib Q}_s=(\frac{2}{3}\pi , \frac{2}{3}\pi)$. 
The same symbols represent equivalent Ni sites. 
(b) Typical stripe solutions within the HF calculation are presented. 
$n_i$'s and $m_i$'s ($i=\bigcirc, \triangle, \square$) are 
the electron occupation number and total spin moment 
(in units of $\mu_{\rm B}$), respectively, at Ni sites marked 
with the same symbols in (a).}
\label{Fig5}
\end{figure}

We consider the following four self-consistent solutions (I)-(IV). 
(I): not allowing lattice distortions, we diagonalize 
$H_{\rm MF}$ from eq.~(\ref{eq:H_MF_2}) and 
use the self-consistency conditions eqs.~(\ref{eq:nq}) and (\ref{eq:mq}). 
(II): allowing lattice distortions, we diagonalize $H_{\rm MF} + H_{\rm l.d.}$ 
from eqs.~(\ref{eq:H_MF_2}) and (\ref{eq:H_ld_2}), and use the self-consistency conditions 
eqs.~(\ref{eq:nq}), (\ref{eq:mq}) and (\ref{eq:uqq}), 
where we use eq.~(\ref{eq:uqq}) only for Ni-O bonds and 
do not use eqs.~(\ref{eq:upp1}) and (\ref{eq:upp2}) for O-O bonds. 
For (II), we take $K_{\rm Ni, O} = 30$ eV. 
(III) and (IV): allowing lattice distortions, we use eq.~(\ref{eq:uqq}) 
only for Ni-O bonds, and use eqs.~($\ref{eq:upp1}$) and ($\ref{eq:upp2}$) 
for nearest-neighbor O-O bonds. 
For (III) and (IV), we take $K_{\rm Ni, O} = 30$ eV and $K_{\rm Ni, O} = 60$ eV, respectively. 
These solutions are summarized in Fig.~\ref{Fig5}. 
Concerning lattice distortions in (II), nearest-neighbor Ni-O bonds shrink 
particularly around Ni sites with excess hole density ($\bigcirc$ sites in Fig.~\ref{Fig5}), 
being qualitatively consistent with previous studies~\cite{Zaanen1994, Kaneshita2008}. 

\begin{figure}
\begin{center}
\includegraphics[width=100mm]{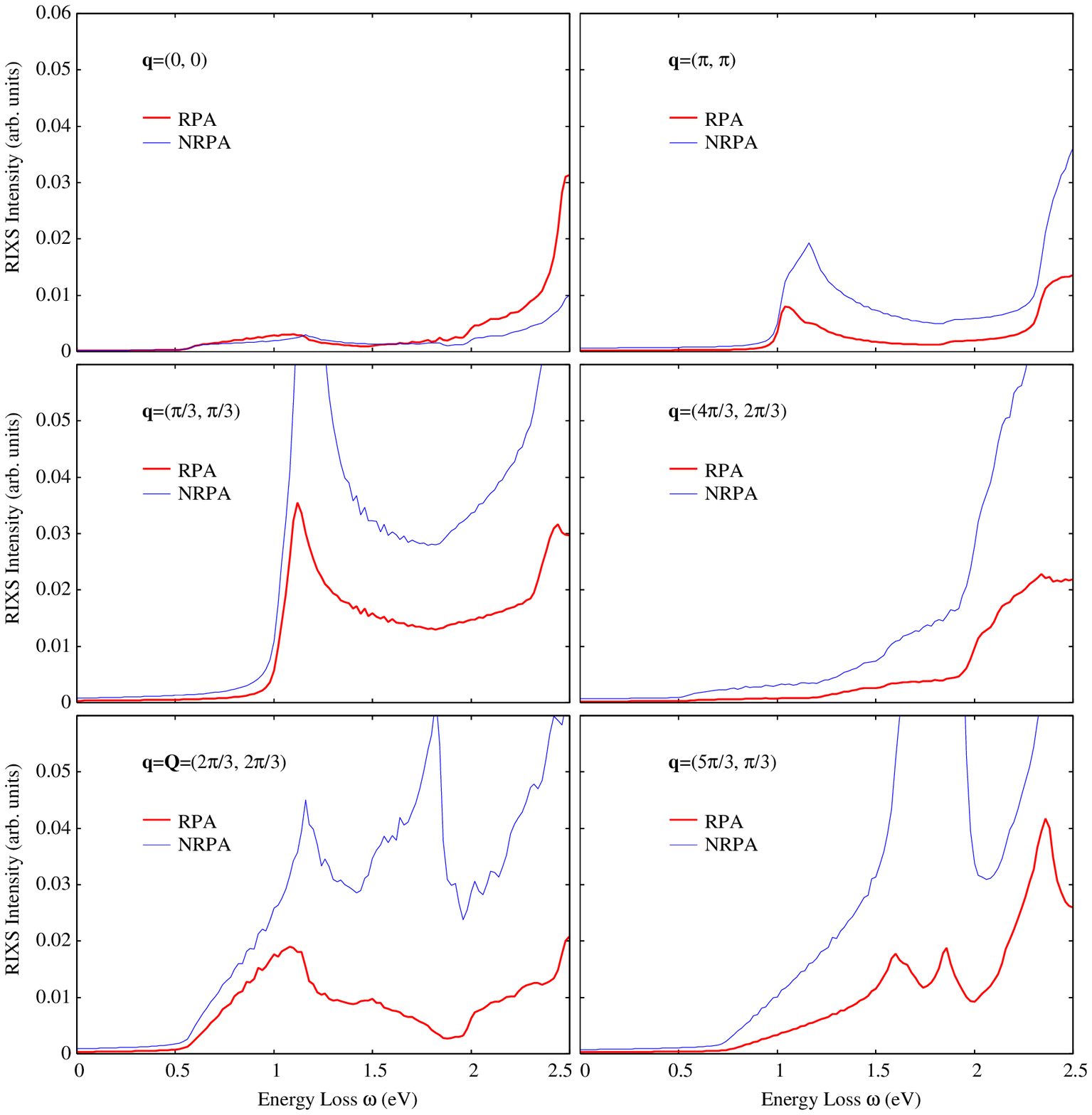}
\end{center}
\caption{
(Color online)
Calculated RIXS spectra in the low-energy region 
at various momentum transfers for the stripe solution (III). 
In each panel, the thick red and thin blue lines represent 
the results calculated with and without RPA, respectively.}
\label{Fig6}
\end{figure}

In Fig.~\ref{Fig6}, we present the calculated RIXS spectra in the low-energy region
at various ${\mib q}$ points on the symmetry lines, for the stripe solution (III). 
A remarkable feature is that the low-energy edge of the spectra shows a tail 
toward the low-energy region at ${\mib q}={\mib Q}=(\frac{2}{3}\pi, \frac{2}{3}\pi)$. 
This is consistent with experimental results by Wakimoto et al.~\cite{Wakimoto2009}. 
It should be noted that this low-energy tail appears already 
in the spectrum calculated without RPA, as shown in Fig.~\ref{Fig6}. 

\begin{figure}
\begin{center}
\includegraphics[width=120mm]{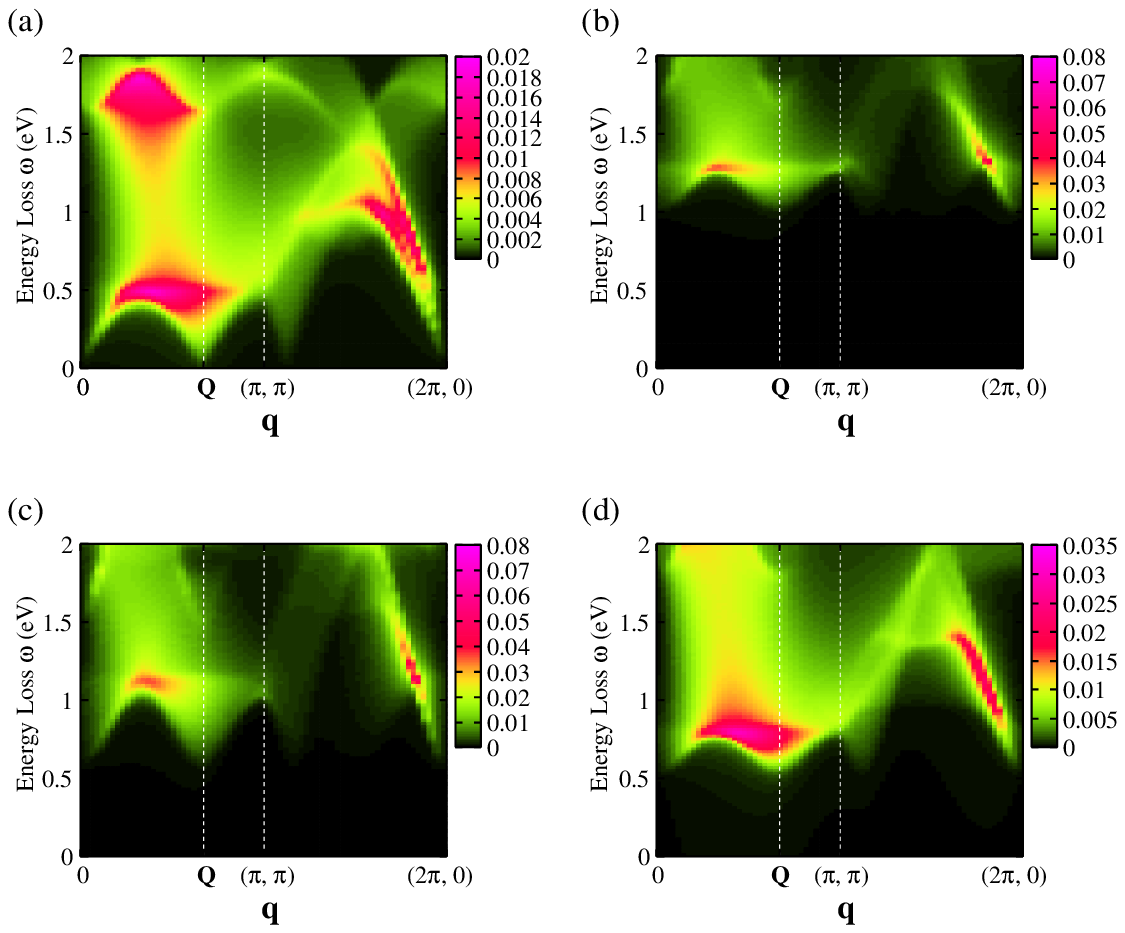}
\end{center}
\caption{
(Color online) 
Panels (a), (b), (c) and (d) show the intensity plots of calculated RIXS spectra 
for the four stripe solutions (I), (II), (III) and (IV), respectively
}  (see the text and Fig.~\ref{Fig5} for the four stripe solutions). 
In each panel, the horizontal and vertical axes represent momentum 
transfer ${\mib q}$ and energy loss $\omega$, respectively, and 
${\mib q}$ sweeps along the diagonals of the square BZ. 
${\mib Q}=(\frac{2}{3}\pi, \frac{2}{3}\pi)$ is the stripe vector.
\label{Fig7}
\end{figure}

To see more detailed momentum dependence of the RIXS spectra on momentum transfer ${\mib q}$, 
we show intensity plots of the calculated RIXS spectra along the diagonal lines 
of the original square BZ for the stripe solutions (I)-(IV) in Fig.~\ref{Fig7}. 
Comparing Fig.~\ref{Fig7} with Fig.~3 of Ref.~\ref{Wakimoto2009}, 
we consider that the RIXS intensity for the stripe state (III) 
is the most consistent with the experimental result. 
Comparing the spectra for the stripe states (II) and (III), 
we can see that shrinkage of nearest-neighbor O-O bonds, 
which is caused by shrinkage of nearest-neighbor Ni-O bonds, 
enhances the low-energy dispersive behavior around ${\mib q} \approx {\mib Q}$. 
Comparing the spectra for the stripe states (III) and (IV), 
we can see that the low-energy gap at ${\mib q} = {\mib Q}$ becomes larger 
for smaller $K_{\rm Ni, O}$, i.e., for softer Ni-O bonds. 

\begin{figure}
\begin{center}
\includegraphics[width=130mm]{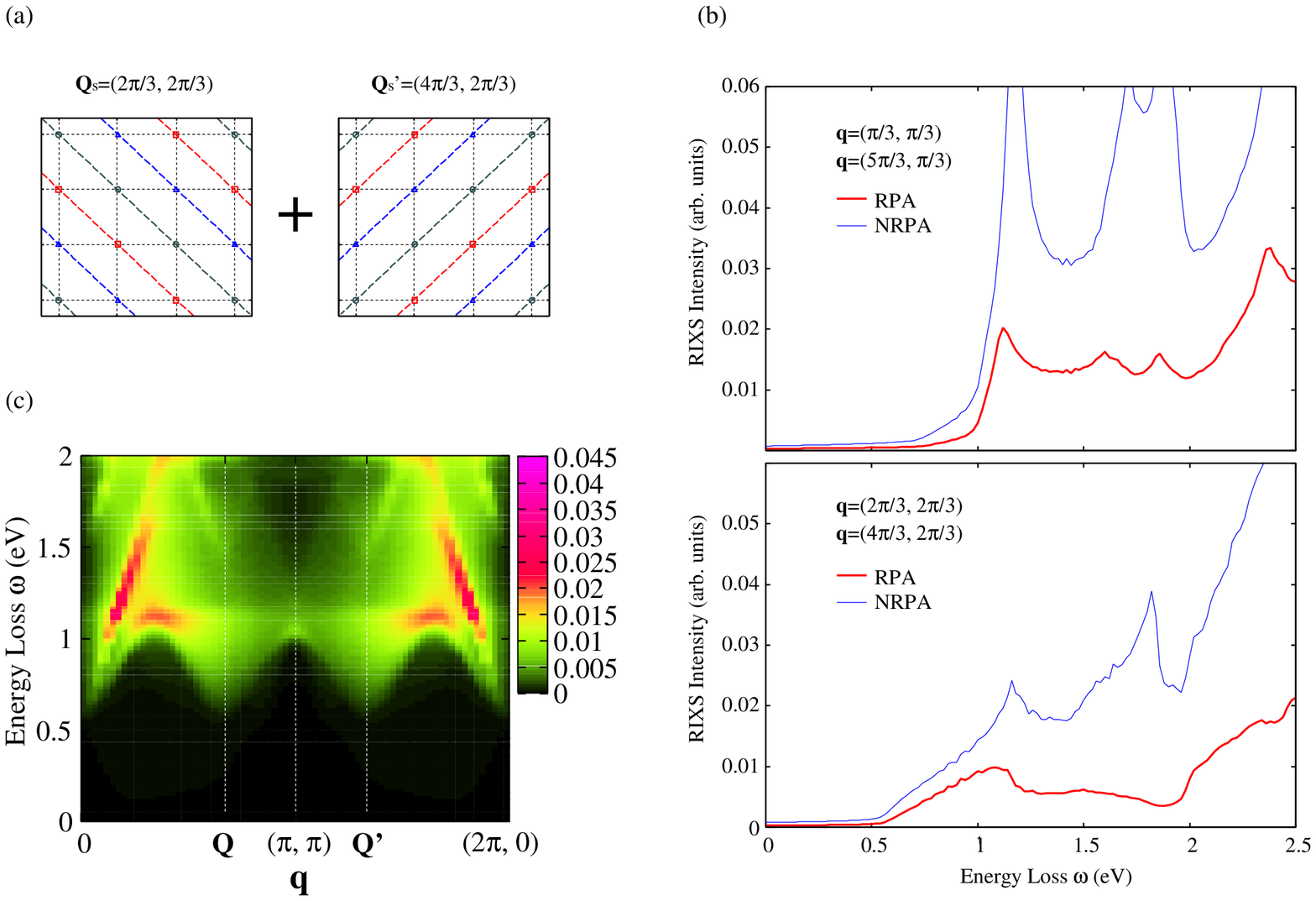}
\end{center}
\caption{
(Color online)
(a) Schematic figure of two kinds of diagonal-stripe domains. 
(b) Averaged RIXS intensity from the two stripe domains 
at ${\mib q}=(\pi/3, \pi/3)$, $(2\pi/3, 2\pi/3)$, 
$(4\pi/3, 2\pi/3)$ and $(5\pi/3, \pi/3)$. 
The thick red and thin blue lines represent 
the results calculated with and without RPA, respectively. 
The spectra at ${\mib q}=(0, 0)$ and $(\pi,\pi)$ 
are unchanged by taking the average, and are presented still 
by the upper two panels of Fig.~\ref{Fig6}.  
(c) Intensity plot of the averaged RIXS spectra. 
The horizontal and vertical axes represent momentum 
transfer ${\mib q}$ and energy loss $\omega$, respectively, 
and ${\mib q}$ sweeps along the diagonals of the square BZ. 
}
\label{Fig8}
\end{figure}

So far, we have considered only a single stripe-ordered domain 
characterized by the stripe vector ${\mib Q}_s=(\frac{2}{3}\pi, \frac{2}{3}\pi)$. 
To compare with the experimental results in more detail, 
we should bear in mind that, in addition to the above kind of domains, 
actual samples will contain also domains which correspond 
to ${\mib Q}_s=(\frac{4}{3}\pi, \frac{2}{3}\pi)$ (see Fig.~\ref{Fig8}(a)), 
as one can expect easily from tetragonal lattice symmetry. 
Here we assume that those two kinds of domains take evenly 
the same volume fraction in the sample, and the domain walls 
separating the domains only negligibly affect the RIXS spectra. 
Under this simple assumption, the total RIXS spectrum is given 
by the average of the contributions from the two kinds of domains. 
The averaged RIXS spectra for the stripe solution (III) 
are presented in Figs.~\ref{Fig8}(b) and (c) 
(compare Fig.~\ref{Fig8}(c) with Fig.~3 of Ref.~\ref{Wakimoto2009}). 

According to the theoretical spectra (the upper two panels of Fig.~\ref{Fig6} 
for ${\mib q}=(0, 0)$ and $(\pi, \pi)$, and the two panels of Fig.~\ref{Fig8}(b) 
for ${\mib q}=(\pi/3, \pi/3)$, ${\mib q}=(2\pi/3, 2\pi/3)$, 
${\mib q}=(4\pi/3, 2\pi/3)$ and $(5\pi/3, \pi/3)$), 
we see that the intensity at ${\mib q}=(0, 0)$ and $(\pi, \pi)$ 
seems relatively weak compared with that at the other momentum transfers. 
One might consider that the weakness of the intensity at ${\mib q}=(0, 0)$ 
is inconsistent with the strength of the experimental intensity 
at the point `A' in Fig. 2 of Ref.~\ref{Wakimoto2009}. 
However, we should note that, in general, the intensity 
around the $\Gamma$ point (i.e. ${\mib q}=(0, 0)$) can be affected easily 
by several realistic factors which are not included in the present model. 
One of them is the effect of disorders in the sample. 
In doped systems, actual samples inevitably contain some disorders 
or disordered domain walls situated randomly, 
while doped carriers are assumed to form ideally periodic configuration 
in the theoretical model. 
In principle, such disordered fractions possessing no characteristic spatial periodicity 
contribute to the intensity at the $\Gamma$ point. 
Another possible factor is the long-range component of Coulomb interaction. 
In doped systems, the long-range Coulomb interaction can change the form 
of the RIXS spectra around the $\Gamma$ point~\cite{Markiewicz2008}. 
In addition, the multiple scattering due to the core-hole potential 
may enhance the spectral intensity around ${\mib q}=(0, 0)$, as mentioned in \S~\ref{Sc:AF}. 
To take account of these effects thoroughly remains an interesting 
but still difficult future work. 

\begin{figure}
\begin{center}
\includegraphics[width=100mm]{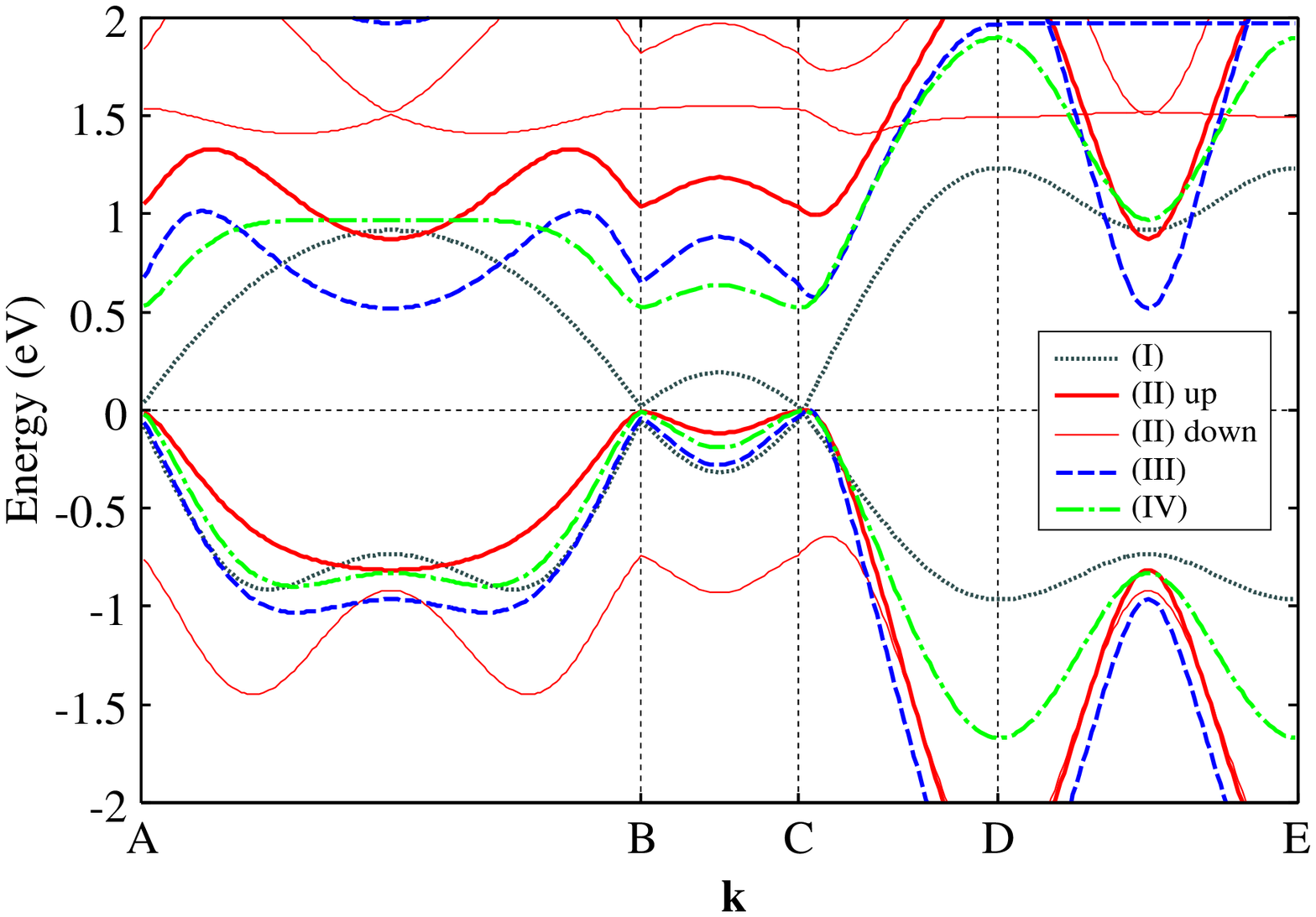}
\end{center}
\caption{
(Color online)
Energy dispersions along the symmetry lines of the reduced BZ (Fig.~\ref{Fig1}) 
in the four stripe states (I)-(IV) (see the text and Fig.~\ref{Fig5}). 
For (I), (III) and (IV), only up (majority) spin bands are depicted, 
while both of the up and down spin bands are depicted for (II). 
Each symbol, A, $\cdots$, E, on the horizontal axis denotes 
the ${\mib k}$ point with the corresponding symbol in Fig.~\ref{Fig1}. 
The Fermi level is set to zero.}
\label{Fig9}
\end{figure}

To study the origin of the above-mentioned anomalous momentum dependence 
of RIXS spectra around ${\mib q} \approx {\mib Q}$, we present 
the electron energy dispersions for the stripe solutions in Fig.~\ref{Fig9}. 
In Fig.~\ref{Fig9}, only the majority (i.e., up) spin bands are depicted 
for the cases (I), (III) and (IV). 
The calculated bands suggest the existence of gap minima around the symmetry 
points A, B and C (and the equivalent points), as seen in Fig.~\ref{Fig9}. 
As easily expected from this, the low-energy tail around ${\mib q} \approx {\mib Q}$ 
in RIXS spectra originates from the charge excitations 
between those symmetry points. 
Such gap minima may be observable by angle-resolved photoemission 
spectroscopy (ARPES). 

\section{Discussions and Conclusions}
\label{Sc:Conclusions}

We have presented the RIXS spectra calculated 
for undoped insulating antiferromagnet 
La$_2$NiO$_4$ and stripe-ordered La$_{5/3}$Sr$_{1/3}$NiO$_4$, 
where we have described the ground state by the HF approximation 
and have taken account of electron correlations within RPA. 
For the undoped case, we have explained the spectral peak positions 
semiquantitatively, and have presented a more detailed plot than in the previous work. 
In the experiment by Collart et al.~\cite{Collart2006}, 
the momentum-transfer dependence along the line $(0, 0)$-$(\pi, 0)$ 
was studied in detail. 
However, our present calculation suggests that the low-energy peak around 4 eV 
shows stronger dispersion along $(0, 0)$-$(\pi, \pi)$ 
rather than along $(0, 0)$-$(\pi, 0)$. 
This contrasts with the case of the insulating cuprate La$_2$CuO$_4$. 
Experimental verification of this suggestion is awaited. 

For doped stripe-ordered La$_{5/3}$Sr$_{1/3}$NiO$_4$, 
we have explained the anomalous momentum-transfer dependence 
of spectra observed experimentally, 
i.e., the calculated RIXS spectra show a tail toward the low-energy region 
when the momentum transfer of photons 
equals stripe vector ${\mib Q}=(\frac{2}{3}\pi, \frac{2}{3}\pi)$, 
being consistent with the recent experimental result 
by Wakimoto et al~\cite{Wakimoto2009}. 
The reason for the low-energy tail at ${\mib Q}$ is 
that the gap minima of electron energy dispersion in the stripe-ordered states 
exist around the symmetry points A$(\frac{14}{9}\pi, \frac{4}{9}\pi)$, 
B$(\frac{4}{9}\pi, \frac{14}{9}\pi)$, C$(\frac{2}{9}\pi, \frac{10}{9}\pi)$ and 
the equivalent points in momentum space. 
This feature of the energy bands in stripe-ordered La$_{5/3}$Sr$_{1/3}$NiO$_4$ 
may be verified by ARPES experiments. 

We searched for other possible self-consistent solutions 
than presented in Fig.~\ref{Fig5} (b), by choosing various initial values 
of $n_{i\ell}$, $m_{i\ell}$ and ${\mib u}_{ij}$ in numerical iterations. 
However, we found only the equivalent solutions, 
which are obtained by rearranging ($n_\bigcirc, m_\bigcirc$), 
($n_\triangle, m_\triangle$) and ($n_\square, m_\square$) 
or by reversing the signs of spin moments. 
The stripe states with lattice distortions ((II), (III) and (IV) in Fig.~\ref{Fig5} (b)) 
are not symmetric under spatial inversion. 
This does not mean that the electron energy and RIXS spectra 
are not symmetric under inversion in momentum space. 
In fact, the obtained electron energy dispersions are symmetric 
(see the dispersions along the D-E line in Fig.~\ref{Fig9}). 
We checked numerically that the RIXS spectra are also symmetric 
under inversion in wave vector space, as a result 
from the symmetry of the electron energy dispersion. 

For the case of stripe ordered states, we have presented theoretical results 
of the RIXS spectra only for the relatively low-energy region 
(0-2.5eV, as in Figs.~\ref{Fig6}, \ref{Fig7} and \ref{Fig8}). 
We should note that the RIXS spectra calculated for doped stripe states 
within the present formulation are reliable only in such low-energy region, 
while those for the undoped antiferromagnetic state are reliable 
up to the high-energy region. 
For the undoped case, the HF approximation presents large magnetic moments 
($m_{i1}=0.78, m_{i2}=0.96$), and consequently the two magnetically split bands 
with an enough large energy gap mimic well the actual Hubbard bands, 
which are separated by the Mott-Hubbard gap of the order of $U$. 
In this case, the electronic structure in the actual antiferromagnetic ground state 
of the Mott insulating phase is well described over a wide energy range 
up to the order of $U$ by the HF approximation. 
On the other hand, in the doped stripe states, the magnetic moments are reduced 
($m_{i1}=0.22$-0.54, $m_{i2}=0.85$-0.88, for the stripe solution (III)), 
compared with the undoped case. 
In this case, the electronic structure only inside the reduced magnetic gap 
can be still described by the HF approximation, although the high-energy Hubbard 
satellite bands are no longer reproduced. 
Very roughly speaking, the magnetic gap for orbital $\ell$ within the HF calculation 
is evaluated to be $\Delta_{i\ell} \sim U m_{i\ell} + J \sum_{\ell'(\neq \ell)} m_{i\ell'} $. 
According to this rough evaluation, $\Delta_{i\ell} \sim 2.6$-7.5 eV for the stripe solution (III), 
while $\Delta_{i\ell} \sim 7.2$-8.5 eV for the undoped antiferromagnetic ground state. 
Thus, for the doped stripe-ordered cases, the calculated RIXS spectra 
are reliable only in the low-energy region up to about 2.6 eV. 
To study RIXS spectra in such doped Mott insulators over a wider energy range, 
we require more advanced methods such as the dynamical mean-field theory 
(DMFT)~\cite{Georges1996}. 

As we have discussed, the bond shrinkage between nearest-neighbor O-O sites 
seems important to obtain that low-energy anomalous dispersion quantitatively 
consistent with the experimental data. 
This is related to the fact that the doped holes occupy mainly the O2$p$ orbitals. 
The RIXS spectra reflect the momentum distribution of the partial component 
of the Ni3$d$ states mixed with the O2$p$ states. 
Therefore, change of the O-O bonding length (or, in other words, 
change of the O2$p$ band width) affects strongly the low-energy RIXS spectra, 
although the contribution of O-O bonding is energetically negligible 
as mentioned in \S~\ref{Sc:Stripe}. 

In conclusion, we would like to point out that the anomalous low-energy 
RIXS weight at the stripe vector ${\mib Q}$ does not indicate some kinds 
of collective charge excitation modes, since it can be explained 
at least qualitatively within simple band-to-band transitions. 
This physical picture seems consistent with the robustness of the charge ordering 
observed under high electric fields~\cite{Yamanouchi1999,Hucker2007}. 

\acknowledgements
The authors would like to thank Prof. Manabu Takahashi for valuable communications. 
E. K. acknowledges financial support from the Grant-in-Aid for Scientific Research 
from the Ministry of Education, Culture, Sports, Science and Technology of Japan.

\end{document}